\documentclass[
aps,%
12pt,%
final,%
notitlepage,%
oneside,%
onecolumn,%
nobibnotes,%
nofootinbib,
superscriptaddress,%
noshowpacs,%
centertags,%
]%
{revtex4}
\RequirePackage{graphicx}

\def\refitem#1{\relax}

\begin{document}

\title{A formula for charmonium suppression}

\author{C.~Pe\~{n}a}
\email{pena@ift.uni.wroc.pl}
\affiliation{Institute for Theoretical Physics, University of Wroc{\l}aw, 
Wroc{\l}aw, Poland}

\author{D.~Blaschke}
\email{blaschke@ift.uni.wroc.pl}
\affiliation{Institute for Theoretical Physics, University of Wroc{\l}aw,
Wroc{\l}aw, Poland}
\affiliation{Bogoliubov  Laboratory of Theoretical Physics, JINR Dubna, 
Russia}

\begin{abstract}
In this work a formula for charmonium suppression obtained by Matsui in 1989 
is analytically generalized for the case of complex $c\bar{c}$ potential 
described by a 3-dimensional and isotropic time-dependent harmonic oscillator 
(THO).
It is suggested that under certain conditions the formula can be
applied to describe $J/\psi$  suppression in heavy-ion collisions
at CERN-SPS, RHIC, and LHC with the advantage of analytical
tractability.
\end{abstract}

\maketitle

\section{Introduction}

The modification of the charmonium production cross section has been studied
using a schematic 3-dimensional harmonic oscillator for the intermediate and
final $c\overline{c}$ pair in \cite{Matsui:1989ig}.
In that reference the distorted wave Born approximation was used for the
two-gluon fusion model and suppression ratios were calculated.
In the present paper, we consider a 3-dimensional THO with a complex and
continuous time dependent frequency.
For such a generalization, we derive the suppression ratio for charmonia 
states and present a formula for J/$\psi$ suppression including feed-down
contributions.

\section{Quantum mechanical evolution of the $c\bar{c}$ state}
The Charmonium suppression ratio was defined as a ratio of two
cross sections by the expression
$S_\psi(t)=\frac{\sigma(2g\to\psi)}{\sigma_0(2g\to\psi)}$ and was
calculated explicitly in Ref.~\cite{Matsui:1989ig}. From Eqs.
(2.22) and (4.17) of that paper the survival probability for the
s-wave can be written in the following form

{\small
\begin{eqnarray}
\label{survival}
S_\psi(t)=\Biggl|\frac{\int_{0}^{\infty }\;dr\ r^2~
\psi(r)\; U_{c\bar{c}} (r,t)}{\lim\limits_{t \to 0}\int_{0}^{\infty }\; dr\ r^2~
\psi(r)\;  U_{c\bar{c}}(r,t)} \Biggl|^2~.
\end{eqnarray}}

\section{Time evolution operator for the THO model}
We make use of the standard path integral approach in order to
calculate the time evolution operator $U_{c\bar{c}}(r,t)$. We
start by considering a 3-dimensional isotropic THO model with the
Hamiltonian {\small $H=\frac{p^2}{2\mu} +
\frac{\mu}{2}~\omega^2(\tau)\ r^{2}(\tau)$}~, where $r$ is the
$c\bar{c}$ separation and the complex function of time
$\omega(\tau)$ enters in the classical equation of motion for the
heavy pair as
\begin{eqnarray}
\label{e1}
\ddot{r}(\tau)+\omega^2(\tau)\ r(\tau)=0~.
\end{eqnarray}

The general solution of equation (\ref{e1}) is a linear
combination given by {\small $r(\tau)= \rho(\tau)
\Bigl(A~\cos{\gamma(\tau)} + B~\sin{\gamma(\tau)}\Bigl)$}, where
{\small $\gamma(\tau)=\int^\tau_{0}\ dt'^\ \frac{1}{\rho^2(t')}$}.
Replacing these definitions into (\ref{e1}), clearly leads  to the
following Ermakov equation \cite{Gjaja:1992}
\begin{eqnarray}
\label{e3}
\ddot{\rho}(\tau)+\omega^2(\tau)\ \rho(\tau) -\frac{1}{\rho^3(\tau)}=0~.
\end{eqnarray}

If $\tau~\in~[0~,~t]$ then A and B can be easily obtained from the
initial conditions as
\begin{eqnarray}
\label{e5}
A=\frac{r(0)}{\rho(0)}~,\hspace{1cm}
B=\frac{1}{\sin \gamma(t)}\left[\frac{r(t)}{\rho(t)}
- \frac{r(0)}{\rho(0)}~\cos \gamma(t)\right]~.
\end{eqnarray}

Where we have used that $\gamma(0)=0$. By replacing A and B in the
general solution, we obtain $r(\tau)$ and $\dot{r}(\tau)$. For a
THO the classical action $s_{cl}$ and the fluctuation factor
$F(t)$  in the 3-dimensional isotropic space are defined in
Ref.~\cite{kleinert}. We calculate here their relationship with
Ermakov function as\footnote{We use the notation
$\dot{r}(t)=\frac{dr(\tau)}{d\tau}|_{\tau=t}$ for all functions of
time.}
\begin{eqnarray}
\label{e8}
s_{cl}&=&\frac{\mu}{2}~\Bigl(
r(t)~\dot{r}(t)-r(0)~\dot{r}(0)\Bigl)\nonumber\\
&=&\frac{\mu}{2} ~\frac{1}{\sin \gamma(t)}\times
\Bigl[r(t)^2~\Bigl(\dot{\gamma}(t)\cos \gamma(t) + \frac{\dot{\rho}(t)}{\rho(t)}~\sin \gamma(t)\Bigl)
\nonumber\\
&&+ r(0)^2~\Bigl(\dot{\gamma}(0)~\cos \gamma(t)-\frac{\dot{\rho}(0)}{\rho(0)}~\sin \gamma(t)\Bigl)
-r(t)~r(0)~\Bigl(\frac{\rho(t)}{\rho(0)}~\dot{\gamma}(t) + \frac{\rho(0)}{\rho(t)}~\dot{\gamma}(0)\Bigl)\Bigl]~,\\
 F(t)&=&\sqrt[3]{\frac{\mu}{2\pi i}\Bigl(-\frac{\partial \dot{r}(t)}{\partial
r(0)}\Bigl)}
=\sqrt[3]{\frac{\mu}{2\pi i}~\frac{\rho(t)~\dot{\gamma}(t)}{\rho(0)
\sin \gamma(t)}}~.
\label{e8.1}
\end{eqnarray}

The time evolution operator for THO is given exactly by
{\small $U(r,t)=F(t)~\exp(i\; s_{cl})$}.
In the present context, it will represent the quantum mechanical evolution of 
a $c\bar{c}$ state for a medium-modified (distorted) interaction up to the 
time $t$ when it gets projected onto the asymptotic bound state spectrum.
Thus we define $U_{c\overline{c}}(r,t)=U(r,t)$.
In fact formula (\ref{survival}) is independent of the initial condition which
may be taken as $r(0)=0$.

\section{The THO formula for charmonium suppression}

The ground state of charmonium $J/\psi$ can be identified with the
1s-wave of the harmonic oscillator given by {\small
$\psi(r)=\psi(0)~\exp\Bigl(\frac{-r^2}{2~r^2_\psi}\Bigl)$} with
{\small $r_\psi=\sqrt{\frac{1}{\mu~\omega_\psi}}$}
\cite{Kopeliovich:2010nw}. Thus we integrate the gaussian shape
over $r$ appearing in (\ref{survival}) which leads to the
following  suppression
\begin{eqnarray}
\label{fsurvival}
S_{J/\psi}(t)&=&
\Bigl|\frac{\rho(t)}{\rho(0)}\Bigl|^3\times
\Bigl|\cos \gamma(t)+\Bigl(\frac{\dot{\rho}(t) \rho(t)^{-1}
}{\dot{\gamma}(t)} + i\;\frac{\omega_\psi}{\dot{\gamma}(t)}\Bigl)~\sin \gamma(t) \Bigl|^{-3}~.
\end{eqnarray}

The formula (\ref{fsurvival}) depends on $\gamma(t)$, the
frequency $\omega_\psi$ and the Ermakov function $\rho(t)$. For
the case of the charmonium state $\psi'$ we take the 2s-wave given
by {\small
$\varphi(r)=\frac{2}{3}~\varphi(0)~\Bigl(\frac{3}{2}~-~\frac{r^2}{r^2_\psi}
\Bigl)~\exp\Bigl(\frac{-r^2}{2~r^2_\psi}\Bigl)$}. Applying the
formula (\ref{survival})  we obtain
\begin{eqnarray}
\label{ffsurvival} S_{\psi'}(t)&=& S_{J/\psi}(t)~\Biggl|~1~-~
\frac{2~i~\omega_\psi~\sin \gamma(t)}{\Bigl(i~\omega_{\psi}+
\frac{\dot{\rho}(t)}{\rho(t)}\Bigl)\sin \gamma(t)+
\dot{\gamma}(t)\cos \gamma(t)}~\Biggl|^2~.
\end{eqnarray}

For the Charmonium state $\chi_c$ we take the 2p-wave  given by
{\small
$\chi(r)=\chi'(0)~r~\exp\Bigl(\frac{-r^2}{2~r^2_\psi}\Bigl)$}.
However, in this case there is a contribution of the angular
momentum and it was shown in Ref.~\cite{Matsui:1989ig} that for
such waves the formula (\ref{survival}) vanishes and the
next-to-leading order term in momentum O(p/m) must be considered
leading to the expression
\begin{eqnarray}
S_{\chi}(t)&=&\Biggl|\frac{\int_{0}^{\infty }\;dr\ r^2~ \chi(r)\;
U'_{c\bar{c}} (r,t)}{\lim\limits_{t \to 0}\int_{0}^{\infty }\; dr\
r^2~ \chi(r)\; U'_{c\bar{c}}(r,t)}
\Biggl|^2=S^\frac{5}{3}_{J/\psi}(t)~, \label{fffsurvival}
\end{eqnarray}
with $U'_{c\bar{c}}=-\frac{\mu~r}{2\sin \gamma(t)} \Bigl(
\rho(t)~\dot{\gamma}(t)~\rho(0)^{-1} +
\rho(0)~\dot{\gamma}(0)~\rho(t)^{-1} \Bigl) U_{c\bar{c}}$~. The
observable $J/\psi$ suppression ratio is influenced by feed-down
from the higher charmonia states and we shall assume the following
composition of the total contribution
\begin{eqnarray}
\label{finalsurvivalno}
S(t)~=~0.6 ~ S_{J/\psi}(t)~ + ~ 0.3 ~ S_{\chi}(t)~ + ~ 0.1 ~ S_{\psi'}(t)~.
\end{eqnarray}

The case of no feed-down is described by the expression
{\small $S_{no}(t)~=~S_{J/\psi}(t)$}~.
Since we have already shown that
{\small $S_{\chi}(t)<S_{J/\psi}(t)$} and
{\small $S_{\psi'}(t)<S_{J/\psi}(t)$}  for
$S_{J/\psi}(t)<1$ it is clear that {\small $S(t)<S_{no}(t)$}.

\section{Summary}

We have generalized Matsui's harmonic oscillator model for charmonium
suppression to the case of time-dependent complex oscillator strengths and
included the effects of feed-down on the $J/\psi$ suppression ratio.
Preliminary results for the comparison with experimental results from CERN SPS
and RHIC can be found in \cite{Blaschke:2011zz}.

\begin{acknowledgments}
The authors acknowledge support from the Polish Ministry for Science and 
Higher Education MNiSW. 
D.B. has been supported in part by the Russian Fund for Basic
Research RFBR under grant No. 11-02-01538-a.
\end{acknowledgments}


\end{document}